# Atomic Imaging Using Secondary Electrons in a Scanning Transmission Electron Microscope : Experimental Observations and Possible Mechanisms

H. Inada,[1,2] D. Su,[1] R.F. Egerton,[3] M. Konno,[2] L. Wu,[1] J. Ciston,[1] J. Wall[1] and Y. Zhu[1*]

[1]Brookhaven National Laboratory, Upton, NY 11973, USA
[2]Hitachi High Technologies Corp., Ibaraki, Japan
[3]University of Alberta, Edmonton, Canada

## Abstract

We report our detailed investigation of high-resolution imaging using secondary electrons (SE) with a subnanometer probe in an aberration-corrected transmission electron microscope, Hitachi HD2700C. This instrument also allows us to acquire the corresponding annular-dark-field (ADF) images simultaneously and separately. We demonstrate that atomic SE imaging is achievable for a wide range of elements, from uranium to carbon. Using the ADF images as a reference, we study the SE image intensity and contrast as a function of applied bias, atomic number, crystal tilt and thickness to shed light on the origin of the unexpected ultrahigh resolution in SE imaging. We have also demonstrated that the SE signal is sensitive to the terminating species at a crystal surface. Possible mechanisms for atomic-scale SE imaging are proposed. The ability to image both the surface and bulk of a sample at atomic scale is unprecedented, and could revolutionize the field of electron microscopy and imaging.

## 1. Introduction

Direct imaging of individual atoms, or their arrangement on the surface and their interaction with the bulk of a sample, has long been recognized as a need as well as a challenge for the microscopy and research community. A modern scanning-tunneling microscope provides atomic resolution, but only of surface atoms of flat specimens. In contrast, a state-of-the-art scanning transmission electron microscope (STEM) can now routinely image atoms in the bulk but has little surface sensitivity, while a modern scanning electron microscope (SEM) can capture surface morphology but does not offer atomic resolution. In the SEM, secondary-electron imaging is the most popular mode of operation and is traditionally used to reveal the sample's surface topography. Nevertheless, this SE imaging method has never been regarded as being on the cutting edge of performance, due to its perceived limited spatial resolution in comparison with its STEM counterpart using transmitted electrons.

The recent achievement of 0.1nm in atomic imaging using the Hitachi HD2700C STEM suggests that SE imaging can now achieve comparable spatial resolution to the STEM mode but with complementary capabilities [1,2]. The electron-optical system enables simultaneous capture of atomic images on the surface (using secondary electrons) and through the bulk of a sample (using transmitted and elastically scattered electrons), thus opening a door for a wide range of applications in both physical sciences and life sciences. An example is shown in Figure 1, where individual uranium atoms (circled) and uranium-atom clusters on a carbon support film are imaged simultaneously with the instrument using a 0.1nm scanning probe, using forward scattering (top right panel) and backward scattering (bottom right panel). The middle right panel of Fig.1 shows a superimposition of the ADF (in red) and SE (green) images. Red spots present in (a) but absent in (b) are presumed to be on the bottom of the substrate, as illustrated in the left panel of Fig.1. Having the ability concurrently to visualize atoms on the surfaces and in the bulk brings a new dimension to materials research, and should allow us one day to determine the active sites of a catalyst during a chemical reaction, for example. Therefore this ultra-high resolution SE phenomenon warrants further understanding and development. In this article we report in some detail on the instrumentation development and various experimental tests of SE imaging, including the study of SE imaging contrast as a function of electrical bias and its dependence on atomic number of a sample, defocus, sample thickness, and crystal tilt. Finally we discuss the possible mechanisms for atomic imaging using secondary electrons.

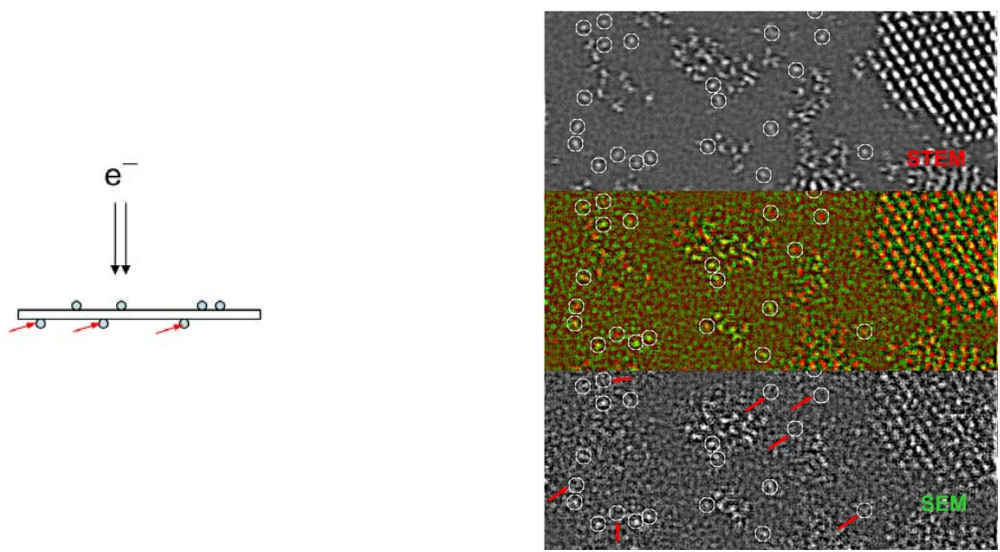

Fig. 1 Simultaneous acquisition of the SEM image using secondary electrons (bottom) and the ADF-STEM image using transmitted electrons (top) of uranium oxide nanocrystals and uranium individual atoms (filtered with unsharp mask in real space) on a 2-nm carbon support. The overlap of the two is shown in the middle. The circled areas with weak signal marked by arrow in the SEM image are interpreted as locations where individual atoms are located at the bottom surface of the carbon support (see the schematic on the left) and secondaries are blocked by the support.

## 2. Instrumentation

The imaging system developed for simultaneous acquisition of SE and ADF image pairs was named "Hitachi HD2700C" right before its delivery to the Center for Functional Nanomaterials, Brookhaven National Laboratory. This instrument is housed in a specially designed laboratory with appreciably minimized mechanical vibration, temperature variation, and electromagnetic field [3,4]. It is the first aberration-corrected electron microscope manufactured by Hitachi and is based on HD2300A, a dedicated STEM developed a few years earlier as an alternative to the discontinued VG STEMs. The BNL instrument has a cold-field-emission electron source with high brightness and small energy spread, ideal for atomically resolved imaging and electron energy-loss spectroscopy (EELS). The microscope has two condenser lenses and an objective lens with a 3.8mm gap, compared to the 5mm-gap objective lens in HD2300A, with the same ±30º sample tilt capability and various holders for heating and cooling (-170~1000°C). The projector system consists of two lenses that provide considerable flexibility in choosing various camera lengths and collection angles for imaging and spectroscopy. The convergence and collection angles for various settings can be found in [3,4].

There are seven fixed and retractable detectors in the microscope. Above the objective lens is the Hitachi secondary electron detector for imaging a sample's surface. Below are the Hitachi analog HAADF (high-angle) and BF (bright-field) detectors for STEM, and a Hitachi TV rate (30frame/sec) 8bit CCD camera (480×480) for fast and low magnification observations and alignment. The Gatan 2.6k×2.6k 14 bit CCD camera located further down is for diffraction (both convergent and parallel illumination) and Ronchigram analysis. The Gatan analog MAADF (medium angle ADF) detector and EELS spectrometer (using a 16bit 100×1340 pixel CCD) are sited at the bottom of the instrument. The spectrometer (Enfina ER) is a high-vacuum compatible high-resolution device that Gatan designed specifically for Brookhaven The CEOS probe corrector, located between the condenser lens and the objective lens, has 2 hexapoles and 5 electromagnetic round lenses, 7 dipoles for alignment, and 1 quadrupole and 1 additional hexapole for astigmatism correction. Other features of the instrument include remote operation, double shielding of the high-tension tank and an anti-vibration system for the field-emission tank. The entire instrument is covered with a telephone-booth-like metal box to reduce acoustic noise and thermal drift.

The Hitachi detector above the sample is positively biased to collect low-energy electrons generated at the surface of the specimen for ultra-high-resolution SE imaging. It is a highly efficient detector with high amplifier gain and low noise, consisting of a Faraday cage, a scintillator, light tubes, and a photomultiplier. In a spherical-aberration-corrected electron microscope, the probe size $d$ (measured as FWHM), as a function of the beam convergence half-angle α, is an incoherent sum of contributions from source size, diffraction limit and chromatic aberration, and is given by $d = \sqrt{\left(\frac{4 i_p}{\beta \pi^2 \alpha_1^2}\right)^2 + \left(\frac{0.6\lambda}{\alpha}\right)^2 + \left(C_c \alpha \frac{\Delta E}{E}\right)^2}$ where $i_p$ is the probe current, $β$ is the source brightness, λ is the electron wavelength at beam energy E, ΔE is the electron-source energy spread, $C_c$ is the chromatic aberration constant of the probe-forming lens. The system spherical aberration (third order) is adjusted to about 0.5μm. The calculated minimum probe size for the instrument is 0.075nm ($C_c$=1.5mm), while the experimentally obtained value using single uranium atoms is <0.1nm. Other major achievable specifications of the instrument include 50nA beam current at 1Å probe size, 0.35eV energy resolution and $7 \times 10^{-7}$ high-tension stability measured at 0.14eV of the zero-loss peak for 60sec.

Traditionally, an electron microscope's spatial resolution is tested using a crystalline specimen and very often the results are limited not only by the lattice spacing of the samples, but also by the thickness of the sample, the deviation from the exact zone axis due to the channeling. The electrostatic potential of a single atom can be considered as a delta function for the purposes of probe measurement. We selected uranium atoms (on a carbon support, <4nm in thickness) for high Z, easy availability and characteristic core-loss spectrum for atomic EELS and STEM. The test specimen is typical of negative staining employed in biological studies except that the uranyl acetate is 100x more dilute. Tobacco mosaic virus (TMV) was included to give a thickness gradient with higher concentration of uranium atoms near the TMV, sometimes forming small clumps. The $UO_2$ species observed on such a sample have a nearest neighbor spacing of 3.4A, but their size, as "seen" by the electrons scattered onto the dark field annular detector, is much smaller.

## 3. Experimental Observations

A schematic of the experimental setup for the simultaneous acquisition of SE and ADF images is shown in Fig.2, using Pd catalyst particles on carbon support as an example. Among the three different imaging modes, SE mode shows the highest contrast of the carbon and a better view of the topological structure of the area. Unsurprisingly, the BF mode, based on the phase contrast, reveals the carbon lattice, while the carbon support is invisible in the ADF mode. Clearly, these images provide complementary structural information.

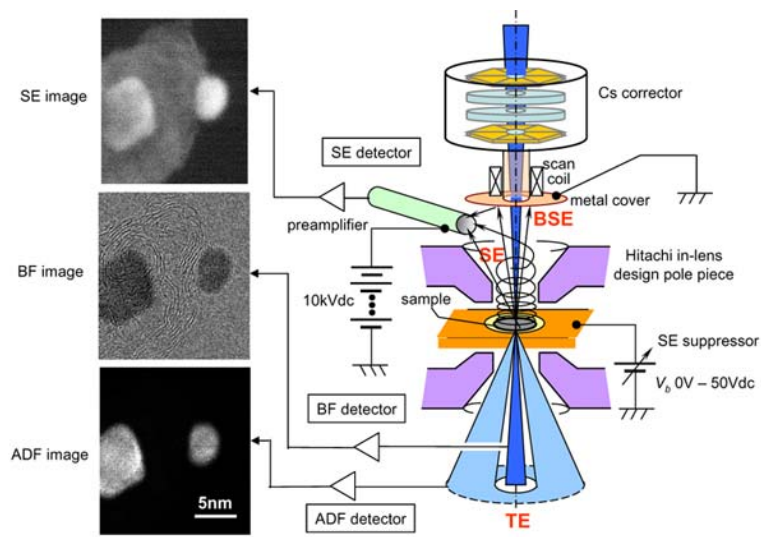

Fig.2 Schematic of the experimental setup for the simultaneous acquisition of SE (image (top left) using secondary electrons (SE) and backscattered electrons (BSE), bright-field (BF) image (middle left) using transmitted electrons (TE) scattered in the forward direction, and annular-dark-field (ADF) image (bottom left) using transmitted electrons that scattered at large angles. The three images are Pd/C catalyst (Pb-core and C-shell) of the same area. We note that BF, ADF and SE images are complementary and that the SE image shows the highest intensity of carbon.

## 3.1 Imaging intensity as a function of electric bias

An important aspect of this study was to understand whether the electrons emerging from the surface are SE or backscattered electrons. One way to separate the two is to apply a bias. Since SE are defined as those that have energies below 50eV, a positive electric bias on the specimen can suppress the emission of the secondaries from the surface, thus the detector can only collect backscattered electrons. The bias experiments were carried out with a modified Faraday Cup sample holder that allows electrical bias to be applied to the specimen.

Fig.3 shows intensity measurements of uranium stained TMV and the underlying carbon film under various biases. Fig.3(a) and (b) are unbiased images, ADF and SE, respectively from the same area. The contrast level of an ADF image, shown in (a), is automatically adjusted in Gatan's DigitalMicrograph software, therefore to quantify the measurement we set the minimum gray level to 0 and maximum to 30,000 electron counts in all of the SEM images. The biased images at 2eV, 5eV, 10eV, 15eV, 25eV and 50eV are shown in Fig.3(c-h). The intensity values were normalized using the corresponding ADF image at 0eV bias as the reference and are tabulated in Table I. The areas A-C are the areas of uranium clusters and D-F are carbon film, as marked in Fig.3(a). The right panel of Fig.3 is the plot of the normalized intensity from both uranium and carbon as a function of bias, revealing a systematic trend of exponential decrease of the intensity as a function of the bias. At a positive bias of 20eV, only 15% signal remains, and can probably be attributed to either high-energy SE or backscattered electrons.

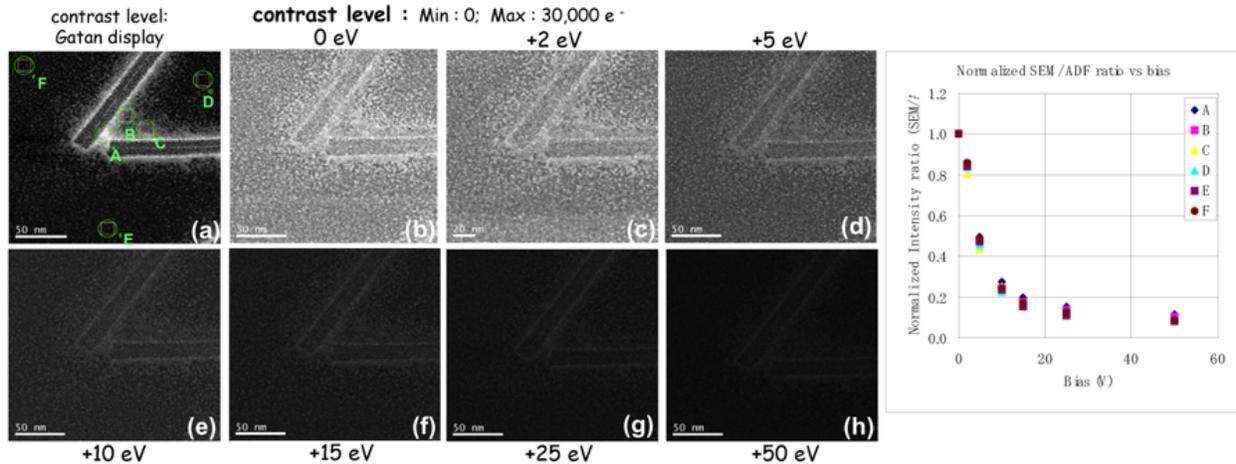

Fig.3 Intensity measurement of U stained TMV clumps and carbon film in (a) ADF-STEM and (b-h) SEM on uranium clusters (A-C) and on carbon films (D-F) as a function of bias: (b) 0V, (c) +2V, (d) +5V, (e) +10V, (f) +15V, (g) +25V and (h) +50V. In commercial electron microscopes, recorded images are automatically converted to a gray scale that is best for display purposes but for quantitative measurement, we set the minimum gray level at 0 and maximum at 30,000 electron counts.

| bias (eV) | uranium | | | carbon | | | standard deviation |
|---|---|---|---|---|---|---|---|
| | A | B | C | D | E | F | |
| 0 | 1.00 | 1.00 | 1.00 | 1.00 | 1.00 | 1.00 | 0 |
| 2 | 0.85 | 0.85 | 0.80 | 0.84 | 0.84 | 0.86 | 0.020 |
| 5 | 0.49 | 0.48 | 0.44 | 0.46 | 0.47 | 0.48 | 0.021 |
| 10 | 0.28 | 0.24 | 0.25 | 0.23 | 0.23 | 0.24 | 0.017 |
| 15 | 0.20 | 0.18 | 0.17 | 0.16 | 0.15 | 0.17 | 0.016 |
| 25 | 0.15 | 0.14 | 0.12 | 0.12 | 0.11 | 0.13 | 0.015 |
| 50 | 0.11 | 0.10 | 0.09 | 0.08 | 0.08 | 0.09 | 0.013 |

Table 1: Tabulated values of the normalized SEM intensities used for Fig.3. The normalization was carried out using the corresponding ADF image at 0eV bias as the reference. STDEV indicates standard deviation of the measurements. The corresponding SEM images at each bias are also shown in Fig.3.

Bias measurements of atomic images are shown in Fig.4, where area A and C are uranium and B and D are carbon and the normalized intensity values are listed in Table II. We note the standard deviation of the measurement at high magnification usually is large compared with the low magnification images, because the areas drift away significantly even under a small bias. Nevertheless, the conclusions are just the same: the majority of signals we detect above the top surface of the samples were from SE rather than backscattered electrons.

The secondary electrons we discuss here are often referred as $SE_I$, i.e., the SE generated by the incident beam (or primary electrons) upon entering the sample. Secondary electrons generated by backscattered electrons when leaving the sample are often referred as $SE_{II}$. Secondary electrons generated by the backscattered electrons striking a lens polepiece or the sample chamber's wall, as well as by primary electrons hitting the aperture are, respectively, referred as $SE_{III}$ and $SE_{IV}$ [5]. Although $SE_{III}$ contain the information of the sample $SE_{IV}$ do not. Furthermore, there are fast secondary electrons, with energy higher than 50eV. Our bias experiments were mainly designed to separate $SE_I$ from backscattered electrons and not from $SE_{II}$, which for the case of a very thin specimen should be negligible. Measurement of other types of secondary electrons, including those with high energy, would require a different bias experiment.

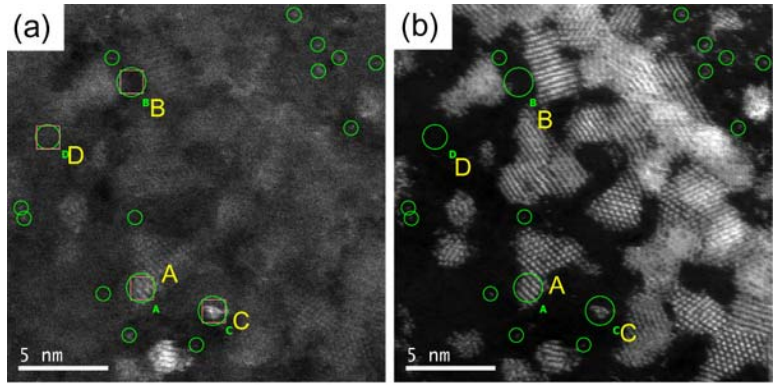

Fig.4 Intensity measurement of uranium oxide on carbon film at atomic resolution in SEM mode (left) and ADF-STEM mode (right) as a function of bias. Only image at 0 bias is shown. Small circles mark spots identified as single atoms.

| bias (eV) | uranium | | | carbon | | |
|---|---|---|---|---|---|---|
| | A | C | STDEV | B | D | STDEV |
| 0 | 1.00 | 1.0 | 0 | 1.00 | 1.0 | 0 |
| 5 | 0.45 | 0.35 | 0.065 | 0.20 | 0.25 | 0.035 |
| 10 | 0.21 | 0.16 | 0.036 | 0.12 | 0.12 | 0.006 |

Table 2: Tabulated values of normalized SEM intensities at high magnification using the corresponding ADF image at 0eV bias as the reference in Fig4. STDEV indicates standard deviation of the measurements.

Another important characteristic of SE is their shallow escape depth, a direct consequence of their low kinetic energy. SE are produced along the primary-electron trajectories within the sample but are subject to elastic and inelastic scattering during their passage through the sample and are strongly attenuated as a result. The probability, $p$, of escape decreases exponentially with depth, $z$, below the surface: $p \sim exp(-z/\lambda)$ where $\lambda$ is the SE mean free path, which depends on the SE energy but is about 1nm for metals and up to 10nm for insulators [5]. To probe the escape depth below the surface, within which SE are generated and escape, a bias experiment was carried out on a 20nm-thick holey carbon film coated with a uniform 2nm thin carbon film; see Fig.5(e) for the sample geometry. Figures 5(a-d) are SE images biased at 0, +3.1, +6.4 and +12.7eV. It is interesting to note that in areas where the 2nm thin film is torn (marked by the ellipses), the underlying holey carbon film is visible. Where the thin film is intact, however, it blocks the majority of the signal from the holey film, i.e. 2nm carbon blocks the SE signal from the holey film underneath. This suggests that our SE signal comes mainly from a region a few nanometers thick below the top surface. The signal from the titanium grid is much less affected by bias, implying it is from the backscattered electrons.

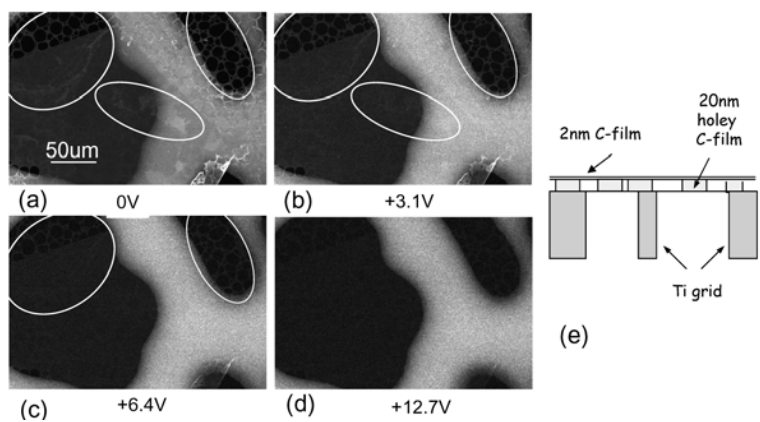

Fig.5  SE/BES images of a thick holey carbon film as a function of bias voltages: (a) 0, (b) 3.1eV, (c) 6.4eV, and (d) 12.7eV. The 20nm thick film is on the top of a Ti grid covered with a continuous carbon film of 2nm thickness (e). The measurement indicates that about 85-90% of the SEM intensity comes from true (SE1) secondary electrons while about 10-15% arises from the backscattered electrons.  Note that where the 2nm thin film is torn (marked by ellipses), the underlying holey film is visible. However, where the thin film is intact, it blocks the signal from the holey film, i.e., the thin carbon blocks the SE signal from the holey film underneath. The signal from the titanium grid is much less affected by bias, suggesting it arises from backscattered electrons.

3.2 Imaging contrast for heavy and light elements (Z=92-6)

To understand the underlying mechanism of SE imaging, we investigated the achievable resolution and image contrast for samples containing elements of different atomic number including U (Z=92), Au (Z=79), Ba (Z=56), Y (Z=39), Sr (Z=38), Cu (Z=29), Ti (Z=22), Si (Z=14), Mg (Z=12), O (Z=8), and C (Z=6).  For most of the experiments, ADF images were simultaneously acquired as a reference.

In our electron microscope, the detection of a single atom in either SE or ADF mode is a result of scanning a 200kV electron beam focused to about 1Å over the specimen.  There is a significant probability that the atom will gain enough energy to remove it from its binding site; hence, a sequence of images will contain information about the movement of individual atoms, limited only by the time resolution of the image acquisition.  This behavior is determined by the balance of several bonding energies, including Van der Waals forces, molecular orbital- and bonding-valence electrons state, surface energy, and electric attraction and repulsion. Fig.6 is an example excerpted from a time series of SE and ADF image pairs, showing the motion of uranium atoms on a 2nm carbon support.  The individual uranium atoms, marked with small circles, are clearly resolved in both SE and ADF images, although the SE images look noisier.  As expected, these individual atoms move around during the observation.  This movement is very similar to that of Au atoms reported by Batson et al [6] and its extent varies from atom to atom, from 1nm to hundreds of nm in a fraction of a second.  Observations on the same sample at liquid-nitrogen temperature do not suggest that the atom motion is reduced.  Since no detectable etching or contamination of the carbon was observed in our sample, these U atoms are believed to sit on the carbon surface and be free to move.  However, some atoms, presumably pinned by local defects, remain stationary on sequential scan lines and from frame to frame.

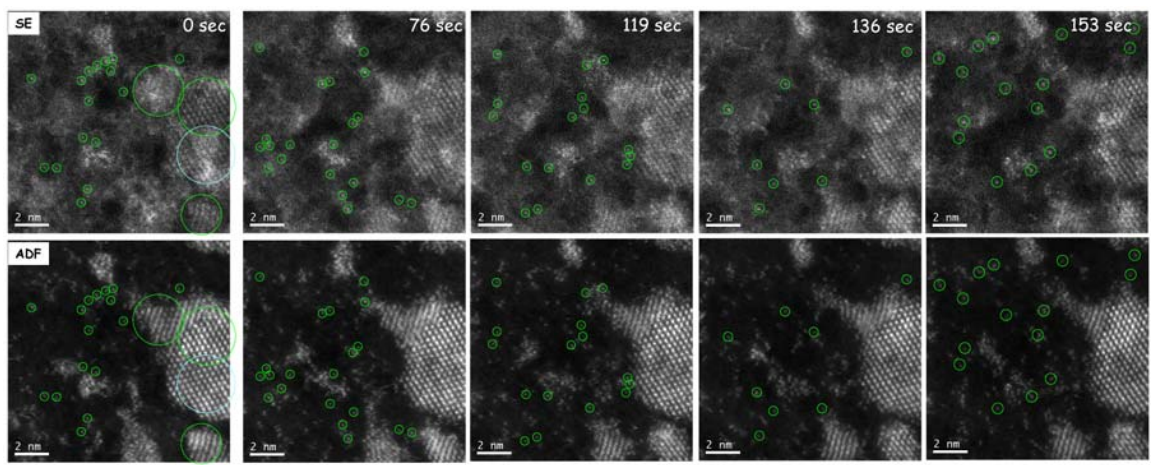

Fig.6 Simultaneous acquisition of SE & ADF image pairs of individual uranium atom and uranium oxide crystals. Image pairs (raw data) were excerpted from a movie to observe atom motion under the 200kV electron beam. (top row) SE images and (bottom row) the corresponding ADF images. The large circles are uranium oxide crystals and the small circles are individual uranium atoms. The corresponding time for each frame is indicated on the SE images. Imaging conditions: Vext=3.57kV Ie=24uA, Ip=210pA, standard mode (1.3A Probe size), α=28mrad and 32.4us/pixel (512x512) @ 8Mx.

Possible delocalization of SE emission was measured by image averaging performed with in-house software (PCMass, available from ftp.stem.bnl.gov) routinely used for mass measurement of unstained biological molecules. Atoms were selected in the ADF image by an automatic algorithm which fitted isolated bright spots to a 2-D Gaussian profile with variable amplitude and width. Spots with the integrated intensity of single U atoms were further aligned one scan line at a time. The intensity of the scan line closest to the atom center and those immediately above and below it were fitted with a 1-D Gaussian profile. Scan lines with amplitude and width profiles typical of heavy atoms were centered and summed with 4x over-sampling (0.00625 nm bin width), using the same offset for ADF and SE traces. This procedure was necessary because of "jitter" from one scan line to the next of roughly one pixel. Test of the alignment procedure on simulated data with the same S/N (56 ADF counts detected per pixel) indicated that the alignment was accurate to within 0.1 pixel. We found that the SE & ADF images are almost equally sharp, with Full Width at Half Maximum (FWHM) of <1A for SE and ADF profiles. The SE curve is noisier due to the low signal level. Fig.7b shows summed intensity profiles of 454 U-atom scan lines in ADF and SE images. As noted previously [1], there is little detectable delocalization (extended tails) in SE images.

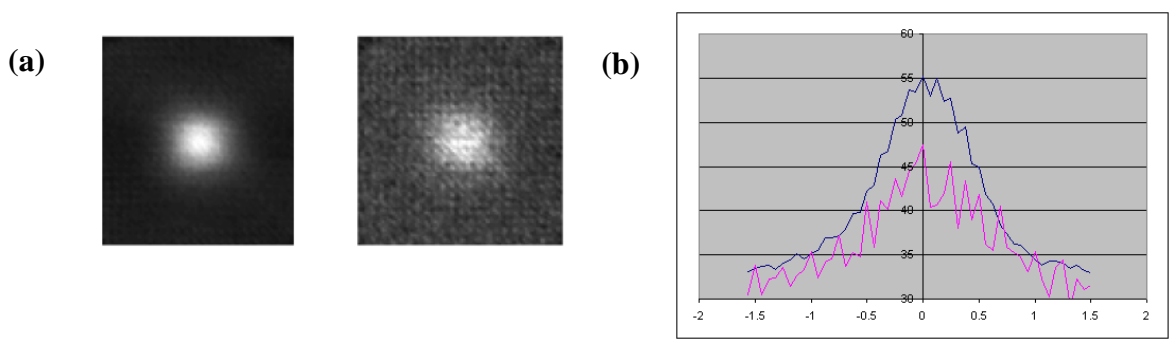

Fig.7 (a) 2-D intensity profiles of individual uranium atoms in the ADF (left) and SE (right) mode. (b) 1-D intensity profiles of individual uranium atoms in the ADF (blue) and SE (pink) mode. There is no detectable delocalization (tails) in atom images.

Another heavy element we examined was Au in the form of nanoparticles, as shown in Fig.8. We note that the ledge in the sample area in the SE image (Fig.8(a)) is considerably more visible than in the ADF image (Fig.8(b)) due to the enhanced signals at the ledge, where the secondary electrons are emitted preferentially from the top and side. Although atomic resolution is achieved in both imaging modes, the SE image has lower contrast with poor counting statistics. We ascribe the noisy SE image to carbon contamination layer on the particles, as seen from the difference in particle size in SE and ADF mode since carbon is invisible in ADF.

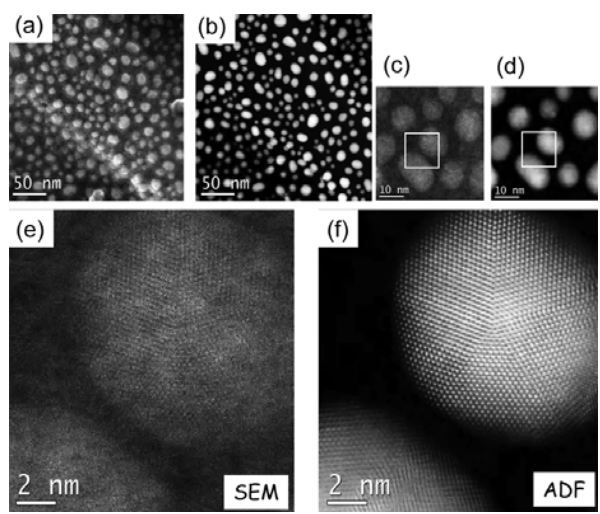

Fig.8 Comparison of SE and STEM-ADF imaging contrast of gold particles on carbon film. (a-b), (c-d) and (e-f) are SE & ADF image pairs with (a, c, e) SE images and (b, d, f) ADF images. Enlarged areas of the Boxed area in (c) and (d) are shown in (e) and (f), respectively. Note the gap between the particles in two imaging modes is very different, due to a thick carbon contamination layer covering the particles, yielding a noisy SE image (also see Fig.9).

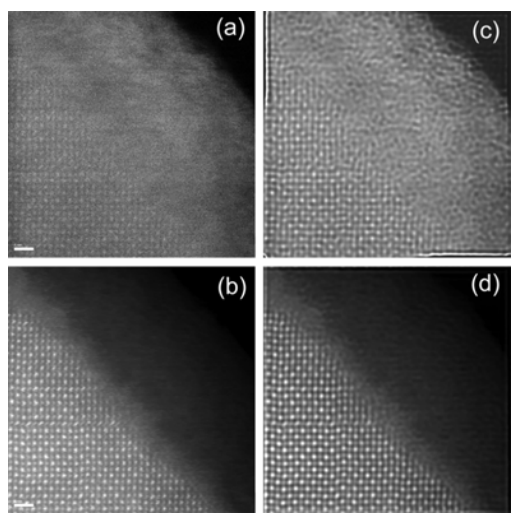

Fig.9 Simultaneous acquisition of SE (a) and STEM-ADF (b) images of a SrTiO$_3$ single crystal (raw data) . (c, d) Maximum-entropy deconvoluted images after 35 iterations of (a, b). A 8nm-thick amorphous-carbon layer grew on the surface during the observation. Note the significant difference in carbon thickness revealed by the SE (top row) and STEM images (bottom row), which is consistent with the different gap-width between the two particles shown in Fig.8(e) and (f).

We also looked at various oxides, including $YBa_2Cu_3O_{7-x}$, $SrTiO_3$ and MgO. In all these samples, atomic resolution of the cations has been achieved in the SE mode. As an example, in Fig.9 we show SE and ADF image pairs of $SrTiO_3$ single crystal, where Fig.9 (a,b) are raw data and (c, d) are deconvoluted images using the maximum entropy method after 35 iterations to reduce background noise. The $SrTiO_3$ sample was a heat-treated single crystal (see details in section 3.3) and the surface amorphous layer was not generated by the ion mill but from carbon contamination during the observation. We note the remarkable difference in the thickness of the carbon layer at the sample edge revealed by SE and ADF imaging, 8nm and 2nm, respectively. It not only implies that SE imaging has much higher sensitivity to low-Z elements than ADF imaging, but also surprisingly suggests that SE can escape from a thick low-Z capping layer if we assume the same thickness of the carbon on the top surface and at its edge of the sample. It is also possible that the signals detected here include both $SE_I$ and $SE_{II}$ as well as backscattered electrons.

Tests on a light element (Si, Z=14) were carried out using pillar samples, prepared with a focused ion beam (FIB) at different ion energies (40kV, 10kV and 2kV) to induce different thickness of the amorphous capping layer of the sample material on the surfaces. Conventional TEM of cross-section samples (Fig.10(a)) reveals that the thickness of the Si amorphous layer induced 40kV ions was 28nm, while 8nm and 3nm for the 10kV and 2kV ions, respectively. We examined the Si samples along the (110) direction using SE and found there is little atomic contrast of the (110) lattice for the sample prepared with 40kV ions. For the 10kV sample, the (110) lattice was visible but the intensity was very low. The atomic images were best seen for the 2kV sample (Fig.10(b)) and in some areas we can clearly separate Si dumbbells at spacing of 0.14nm with 15% contrast (Fig.10(d)). To our knowledge, this is the highest spatial resolution ever achieved in SE imaging. The speckle contrast in Fig.10(b) suggests that the thickness of the amorphous layer may not be even. A Fourier Transform of Fig10(c) indicates that the high spatial frequency of 0.1nm, corresponding to $d_{115}=105$pm, has been transferred. The high spatial resolution SE signals detected from a Si crystal covered with a 3nm capping layer is in contrast with the case for carbon (Fig.5) where a 2nm carbon film almost blocks the emission of the holey carbon film underneath. Apparently the thickness of amorphous layer on the top of a crystal that allows for SE emission also depends on the atomic number of the crystal as well as that of the capping layers.

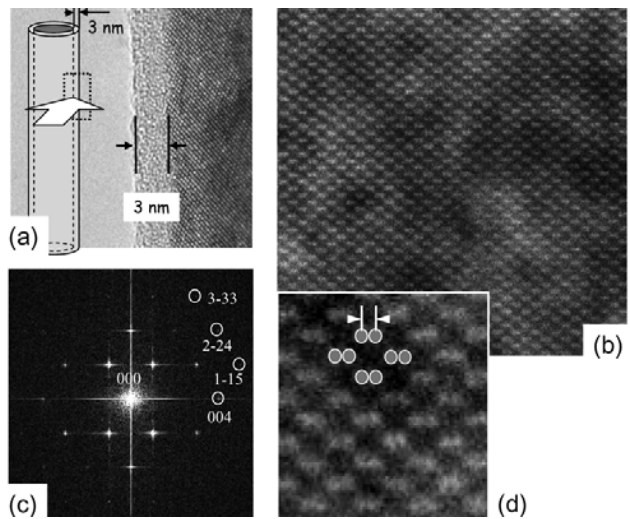

Fig.10 SE images of Si covered with a 3nm-thick amorphous layer on a pillar shaped sample prepared by FIB at 2kV. (a) The schematic of the sample geometry and a conventional HRTEM image providing a cross-section view of the sample that is coated with the amorphous layer. (b) High-resolution secondary electron (SE) image of the sample covered with a 3nm amorphous layer., with speckle contrast due to variations in thickness of the amorphous layer. (c) FFT of the image in (b), showing that high

frequency corresponding to 0.1nm has been transmitted (d004 = 136 pm, d $_{2\text{-}24}$ = 111 pm, d $_{3\text{-}33}$ = 105 pm, and d $_{1\text{-}15}$ = 105 pm). (d) an enlarged area of (b) showing Si dumbbells at 15% contrast.

We also investigated carbon (Z=6) using SE, starting with carbon nanotubes. Imaging carbon nanotubes with SE is very challenging since we usually acquire an ADF image first from the area of interest for local aberration correction but, due to the small scattering cross-section of carbon, good ADF images are difficult to obtain. Furthermore, nanotubes are quickly damaged under a focused 200kV electron beam and contamination is another important issue. Thus, our attempt to image carbon nanotubes was not successful. Nevertheless, we managed to record lattice images of carbon graphite as well as core-shell structures with Pt catalyst particles as the core and carbon as the shell. Fig.11(a-c) is an example showing the same area imaged with BF, ADF and SE mode. Remarkably, the 3.4A lattice spacing of the carbon shells are clearly visible in the SE images. As expected, only the Pt particles can be seen in the ADF image and the carbon shells are not visible. Although the BF image clearly shows the 3.4Å lattice fringes of the carbon, the lattice image looks very flat due to the projection nature of BF imaging. In contrast, the SE image of the same area provides a good topography view of the area. The top layer of the carbon shells is also visible (the intensity in the center of the shells is higher than the nearby vacuum), as shown in Fig.11(c), but is completely invisible in Fig.11(a) and (b). The improved 3-demensional visualization by SE imaging is clearly very important, especially for catalysis research. The 3-D visualization gives a very valuable additional mental clue, particularly whether the heavy particles (catalyst) are inside or outside of their support, towards understanding their catalytic behavior [7].

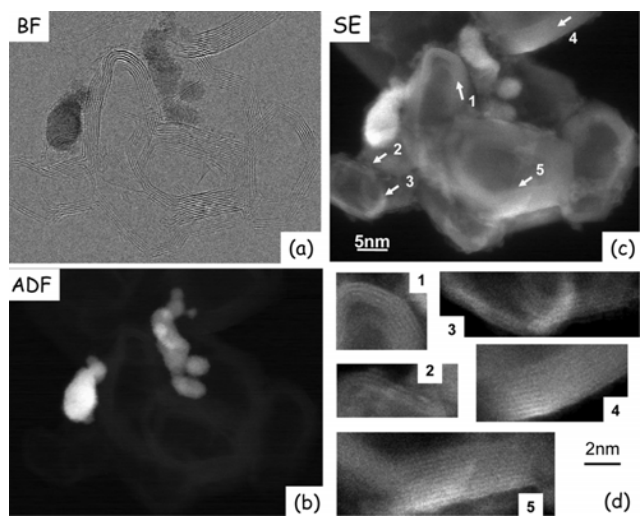

Fig.11 Image contrast comparison of core-shell catalyst particles (Pt-core and C-shell) in three different STEM modes: BF (a), ADF (b) and SE (c) recorded at 200kV.The 0.34nm carbon lattice-fringes are best in the BF image (due to the phase interference) and poorest in the ADF image, while clearly visible in the SE image. (d) Enlarged areas of (c), numbered 1-5. Note the three-dimensional appearance of the SE image in comparison with the other two.

3.3 Image contrast as a function of sample thickness

During our study of SE imaging we soon realized the sample's surface condition is crucial to obtain good data, as in the case of any surface-science experiment. We thus tried different ways to improve surface quality by heat-treating the samples. MgO and SrTiO$_3$ single crystals were cut to 3mm discs, mechanically dimpled to <5um at the center, then thinned using a conventional ion-mill with Ar$^+$ at low energy (<5kv). The self-supporting single crystals were then annealed to re-crystallize surface damage from sample preparation, and to produce atomically flat crystalline surfaces. MgO crystals were annealed in air at 800C for 3 hours followed by 1000C for an additional 3 hours to produce the $\sqrt{3}\times\sqrt{3}$-R30 surface reconstruction according to [8], which was confirmed by parallel beam electron diffraction. Nevertheless, the high-resolution imaging of ADF and SE was not successful for MgO, largely due to the charging of the sample. Further study to eliminate the charging problem is under way.

$SrTiO_3$ was annealed for 3 hours at 1000C in air, but surface reconstruction of heat-treated $SrTiO_3$ crystals was not observed, though 1x1 surface reflections were observed in parallel beam diffraction indicating a well-ordered surface of bulk periodicity. ADF imaging of the edges appeared to be clean with negligible damage, making it more difficult to find amorphous areas suitable to tune the aberration corrector. These crystals also possess distinct (001) facets and steps on the surface (Fig.12(a)), ideal for examining the SE image intensities as a function of sample thickness. Figures 12(b-f) are SE (top row) and ADF (bottom row) image pairs for $t/\lambda=0.25\sim 6$, where t is the thickness measured by EELS and $\lambda$ is the inelastic mean free path. For $SrTiO_3$, $\lambda$ is about 115nm for 200kV electrons [9].

We first noticed that atomic images for both SE and ADF can be obtained from relatively thick areas, such as the one shown in Fig.12(f) where t=690nm ($t/\lambda=6$). Though not surprising for the SE image, it is remarkable for the ADF image. However, we observe that the ADF lattice-image periodicity from this thick area does not represent the periodicity of the crystal potential (see the calculated ADF image in (Fig.12(f), bottom)). Such an image artifact does not occur for the corresponding SE image (Fig.12(f), top), which is very similar to the image from the thin area ($t/\lambda=0.25$ (Fig.12 (b)). Straightforward image interpretation may be another advantage of SE imaging.

We measured image contrast in both real space and reciprocal space. In real space, contrast is defined as $(I_{max}-I_{min})/2(I_{max}+I_{min})$ where $I_{max}$ and $I_{min}$ are the maximum and minimum intensity of on and off the atom positions, while in reciprocal space the contrast can be taken as $I_{110}/I_{000}$ where the intensity ratio of the 110 Bragg spot (or other reflection) and the center spot is measured from the Fourier transform of the real-space image. Using both methods, we observe that SE image contrast is much less sensitive to sample thickness than ADF contrast, which exhibits a clear trend of non-linear contrast reduction with increasing crystal thickness, in good agreement with the calculations reported in [10,11]. It is interesting to note that the ADF contrast maximum does not occur at the thinnest area but around $t/\lambda\sim 0.9$ (Fig.12(c)). We attribute this observation to the existence of a surface amorphous layer. In very thin regions, the volume fraction of the crystal is small in comparison with that of surface amorphous, yielding low image contrast. This non-linear contrast reduction with thickness in ADF images is much less noticeable in SE images and the contrast remains almost flat within measurement error over the thickness range of Fig.12.

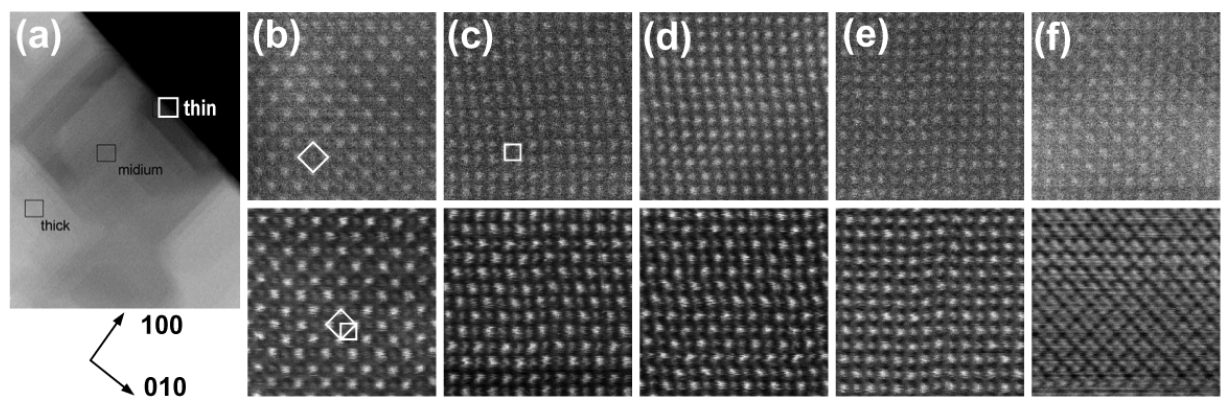

Fig.12 Simultaneous acquisition of SE and STEM-ADF image from a heat treated STiO₃ single crystal as a function of crystal thickness. (a) STEM-BF image of the crystal showing distinct crystallographic (001) facets and steps. (b-f) SE images (top row) and the corresponding ADF images (bottom row) for different t/λ, where t is the thickness and λ is the mean free path (~115nm for SrTiO3): t/λ=0.25 (b); 0.5 (c); 0.9 (d); 1.2 (e) and 6 (f). ADF collection angle : 46~104 mrad.

Careful inspection of the SE images in Fig.12(b-f) suggests that there are two different image patterns, one is a simple square lattice with equal intensity of atoms at the corner and a lattice parameter of 0.38nm (Fig.12(c, d and e)), as illustrated by the square in Fig.12(b). The other is also a square lattice, but body centered and rotated 45° with a lattice parameter of 0.54nm (Fig.12(b and f)), as illustrated by the square in Fig.12(c)). In SrTiO$_3$ only basal-plane O lattices, alternately stacking along the [001] direction, have such an arrangement. It is possible that we are seeing oxygen atoms here rather than Sr and Ti atoms, oxygen being very good at lowering the surface work function, as seen in field emission tip studies. Nevertheless, the mean energy loss per atom is 100eV for oxygen, 220eV for Ti and 370eV for Sr [12] so it is unlikely that O atoms contribute more to secondary emission than Sr and Ti. In addition, the corresponding ADF images do not indicate that the white dots in the SE images are O sites.

A likely scenario, based on the white dots being Sr and Ti atoms, is that the SE images in Fig. 12 (b and f) represent the surface layer with SrO termination, while those in Fig.12 (c, d, and e) have TiO$_2$ termination. Since surface atoms contribute more to SE emission than those embedded beneath the surface and the contribution has an exponential falloff with the distance from the surface (see section 3.1). When the SrO layer is on the top, the Ti atoms are directly underneath a layer of O atoms, and the O/Ti atoms in the column are barely visible, shown as weak dots in the center of the Sr square lattice. When the TiO$_2$ layer is on the top and SrO layer is below (which for a single TiO$_2$ overlayer, the Sr atoms are not directly covered by either O or Ti), the contribution of the Ti atoms to the SE emission is comparable to that of Sr, thus both Ti and Sr atoms have similar intensity. Consequently, a body-centered square (a=5.4A) becomes a simple square (a=3.8A) with a 45° rotation. This is similar to the order-disorder transition of a Perovskite structure in which atoms on the A-site and B-site become indistinguishable. It implies that SE imaging has the potential to image the surface termination layer of a crystal.

3.4 Image contrast as a function of crystal tilt

In order to further investigate the depth dependence of the SE signal, we measured the image contrast as a function of crystal tilt from 0-4 degree (0-67mrad) in both SE (Fig.13, top row) and STEM (Fig.13, bottom row) modes. We note that in ADF imaging, the contrast drops faster with tilt angle than that in SE imaging (especially within 1-2°), indicative of a stronger channeling effect in the ADF mode. The effect of crystal tilt on ADF imaging has been investigated using image simulations of Si crystals [9], where it was reported that the Si 110 lattice can appear to be "on axis" even if it is actually over 2/3deg (12 mrad) off-axis. Nevertheless, a small crystal tilt, even on the order of 0-15mrad, can reduce the contrast of high-resolution ADF images of the crystal by as much as a factor of 2, being consistent with what we observe in the ADF image in Fig.13(b) (bottom) for 1° (17mrad) tilt. This effect of small crystal tilt on ADF images is due to dynamical scattering. The outcome of a small tilt is comparable to lattice displacement, or strain, in a crystal where atoms are misaligned along the optical axis, resulting in a broadened and reduced channeling peak for the incident probe [13]. The difference in atomic misalignment between strain and tilt is that the latter also depends on sample thickness, more misalignment for atoms further down in depth from the surface where the probe is focused. Since SE imaging is surface sensitive, i.e., only a limited number of atomic layers contribute to the SE contrast, the atomic misalignment due to the tilt is much smaller. As we see in Fig.13(a) and (b), for the 1° tilt there is little noticeable change in SE image contrast as well as in the corresponding diffractograms. Significant contrast reduction in SE imaging starts beyond 2° tilt. Quantitative comparison of the image contrast between SE and STEM images of Fig.13 is difficult due to the sudden increase in contrast of STEM-ADF images in the direction perpendicular to the tilt axis at 2-3° arising from the strong two-beam excitation. At small tilts (<1°), the contrast reduction rate for ADF is higher ( ≥5%) than for SE imaging.

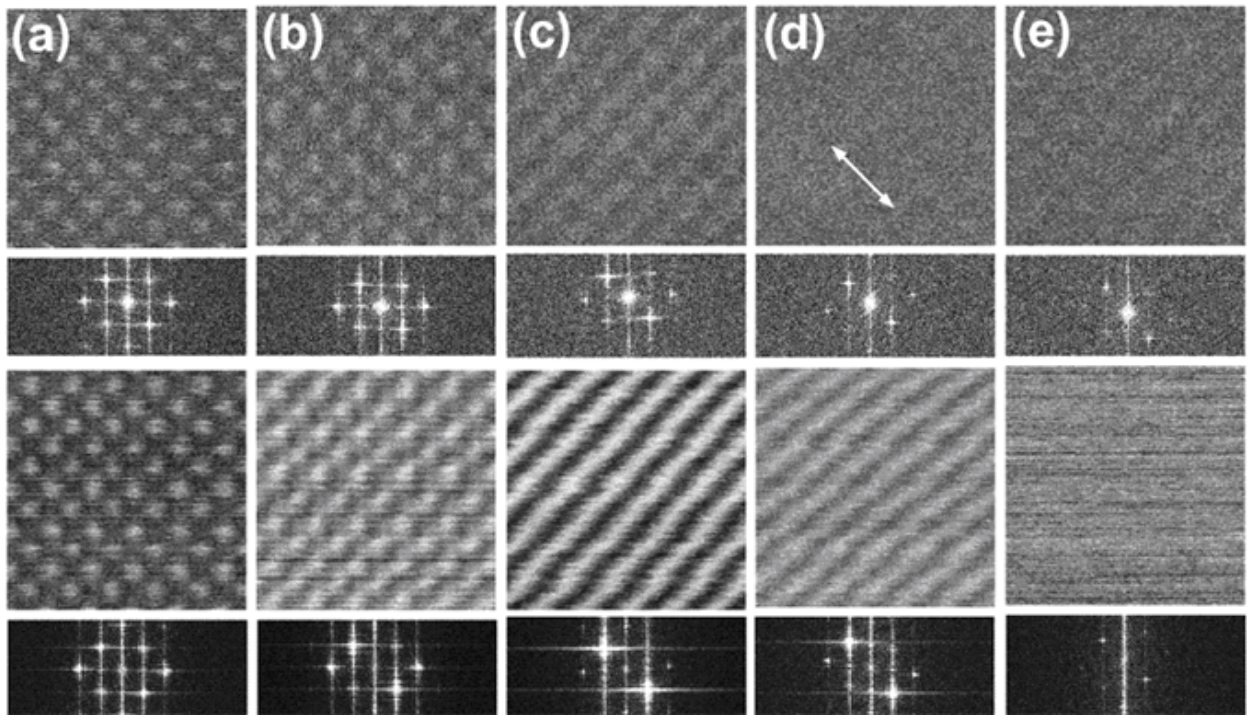

Fig.13 SE and STEM-ADF image pairs of SrTiO3 (100) as a function of crystal tilt. Top two rows: SE images with their Fourier transform shown below; Bottom two rows: the corresponding STEM images and their Fourier transform. (a) 0°; (b) 1°; (c) 2°; (d) 3°; and (e) 4°. The tilt direction, as marked in (d) is near the [110]. The sample thickness is 15nm (t/λ=0.13).

3.5 Image contrast as a function of defocus

We also examined image contrast as a function of defocus for both SE and ADF imaging modes. From pairs of high resolution images alone, the exact focus value of the objective lens appears to be difficult to define. For our Hitachi HD2700C, the depth of focus for ADF is about 5nm at a probe size of ~1Å  We find that if a sample has a 15nm thick amorphous layer on the top of a crystalline structure, a similar amount (15nm) defocus is needed to obtain high quality atomic-resolution ADF images [11].

Since we observe the focus value for best image of SE and ADF is usually the same, we assume the probe is focused on the top layer of the crystal, which may not be the surface of the sample. We thus select that focus position as the "in-focus" condition in our experiment. We vary the lens excitation to alter the focus value using the minimum focus step available (6nm/step) on the instrument.

Fig.14 shows seven pairs of $SrTiO_3$ high-resolution images in the SE and ADF mode. These images were acquired at a convergence angle of 28mrad and defocus up to ±36nm (±6 steps) was used. As we clearly see in the figure, while the in-focus position of the two imaging modes is about the same, the focal depth for SE images is shorter than for ADF. The contrast reduction plot as a function of defocus has a bell shape and is rather symmetric across the "in-focus" position for both imaging modes. At a defocus value of 12nm, although the 0.4nm lattice in SE image (Fig.14(c,e), top row) is still visible, the contrast drops more than half, while for the ADF image (Fig.14(c,e) bottom row) the contrast drops less than 20%. From a simple geometrical consideration, we know for a semi-angle of $\alpha$ = 28mrad and $\Delta z$ = 12nm defocus, the focal spread is $\alpha \cdot \Delta z$ = 0.67nm. This value is not far from what we observe in Fig.14(c,e) and is consistent with a SE escape depth of no more than a few nm.

Unlike STEM-ADF images, which can now be quantitatively described by coherent propagation of wave scattering using a multislice algorithm based on frozen phonon approximation [14], the theory of SE image simulation has not been developed. However, we know that the secondary electrons that contribute to the final SE image are incoherent and have a substantial spread in energy. A simple approximation for quantitative analysis is to convolve the probe intensity with an inelastic point-spread function for each atom and multiply by an escape function for each depth below the surface. This concept is consistent with our defocus experiment, which suggests that probe geometry is an important factor in determining the spatial resolution and contrast in SE imaging.

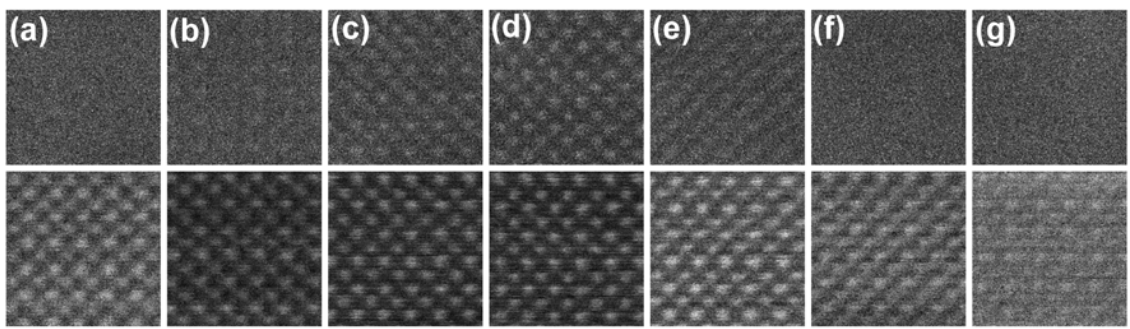

Fig.14 Comparison of image contrast in SE (top row) and STEM-ADF (bottom row) mode as a function of defocus. (a-g) Image pairs taken at seven focal steps with one step corresponds to 6nm of defocus. (d) shows the best contrast for both imaging modes, thus defined as in-defocus. (a-c): under focus, with the defocus value of -36nm, -24nm, -12nm, respectively. (e-g): over focus, with the defocus value of +12nm, +24nm and 36nm, respectively.

## 4. Mechanisms of atomic-scale contrast in secondary-electron images

Quantitative analysis of SE imaging is still in its infancy compared with its counterpart of STEM imaging [15, 16]. The spatial resolution of SE imaging in dedicated STEM instruments and the delocalization of SE signals were studied in the past [17-26]. STEM imaging of small catalyst particles in light atom supports with around 1nm spatial resolution using the SE signal caused some surprise about 20 years ago [7]. It was "explained" on the argument that, even if the SE travel quite long distances before escaping from the specimen, the image resolution is really determined by the incident probe function of energy $E_0$ and by the localization of the process in the initial SE generation step. Although some SE can be produced as much as 1 nm from the probe position by direct excitation of valence electrons, there is a substantial additional or even dominant component arising from excitation of inner-shell electrons, whose mean energy loss *above threshold* is higher than for the valence electrons, resulting in better localization.

The processes that generate contrast in secondary-electron images have been summarized by Reimer [26] and the attainment of nm-range resolution in thin specimens is described by Howie [27] and Liu and Cowley [20, 21]. In general, there are four steps involved in the production of the signal that is used to form a secondary-electron image:
(1) the generation of secondaries through the inelastic scattering of primary electrons in the specimen, the generation rate being G;
(2) random motion of these secondaries, which are scattered by atoms of the specimen both elastically and inelastically (potentially creating other secondaries of lower energy), such that T secondary electrons reach the specimen surface;
(3) the escape of secondaries over the potential barrier at the surface of the specimen, with an average probability P;
(4) acceleration of the emitted electrons in vacuum, such that a fraction D reach the electron detector.

The secondary signal S is a product of these four factors: S = G.T.P.D, and for D = 1 the secondary-electron yield is: $\delta$ = G.T.P = $S/I_0$ where both the probe current $I_0$ and the secondary signal S are expressed in terms of numbers of electrons. It should be emphasized that the escape probability involved in step (2) depends strongly on the depth *z* at which a secondary is generated, and becomes small for secondaries generated below the so-called escape depth. Consequently the term T represents a sum or average over depth *z*. The situation is further complicated in the case of thick specimens, where some (type-2) secondaries are generated by backscattered primaries.

To generate contrast in a scanned-probe image, one or more of the above steps must be depend on the *x* –coordinate of the electron probe in the scan direction, *i.e.*

$$dS/dx = I_0\, T\, PD\, (dG/dx) + I_0\, G\, BD\, (dT/dx) + I_0\, G\, TD\, (dP/dx) + I_0\, G\, TP\, (dD/dx) \qquad (1)$$

In the case of most SE images obtained in an SEM, dT/dx provides the main contrast mechanism: secondaries created at an inclined surface or close to a surface step have an increased probability of escape, resulting in surface-topography contrast [26]. Less commonly, variations in surface work function contribute additional contrast by providing a non-zero dP/dx. In voltage-contrast applications, changes in surface voltage provide a non-zero dD/dx. Atomic-number contrast is possible if the specimen is chemically inhomogeneous and G varies with atomic number.

The case of thin specimens in a STEM is similar to the above, but if we restrict our interest to atomic-scale contrast, we can assume that dT/dx, dP/dx and dD/dx are small. This assumption is reasonable because the scattering process (2) disperses secondaries over a range of *x* that is comparable to the escape depth, typically 1 – 2 nm. Consequently, G, P and D are *x*-averages that vary little with *x* on an atomic scale. For uranium atoms on a carbon substrate, the argument is even simpler: these atoms lie outside the solid, so the terms T and P are not applicable.

This leaves dG/dx as a possible source of atomic-scale contrast. Because secondaries are generated through inelastic scattering of the primary electrons, dG/dx is limited by what is usually called delocalization of the scattering process. Basically this means that a single atom, if imaged using inelastically scattered primary electrons, would have a delocalization width $L(E)$ that is a function of the energy loss *E* involved in the inelastic scattering. In crystalline specimens, delocalization is complicated by simultaneous elastic scattering and has been the subject of numerous calculations [28, 29]. But within a thickness as small as the SE escape depth, elastic scattering is weak and delocalization should be related only to the Lorentzian angular distribution of inelastic scattering. Comparison with experimental data suggests [30] that $L(E)$ can be approximated by a Rayleigh-type formula:

$$L(E) \sim 0.6\, \lambda/\langle\theta\rangle \sim 0.5\, \lambda/(\theta_E)^{3/4} \sim 0.8\, \lambda\, (E_0/E)^{3/4} \qquad (2)$$

where $E_0$ and $\lambda$ are the primary-electron energy and wavelength, $\langle\theta\rangle$ and $\theta_E \sim 0.5 E/E_0$ are median and characteristic angles for inelastic scattering.

Secondary electrons are generated by inelastic collisions that involve a wide range of energy loss and whose relative probabilities are represented by an energy-differential cross section $d\sigma/dE$ or by an energy-loss spectrum recorded with a large collection angle. To simplify the situation, we can replace E in Eq.(2) by a mean energy loss, given approximately [31] by $E_m \sim 6.75 Z$ where Z is the atomic number of the scattering atom. For a uranium atom and $E_0 = 200$ keV, $E_m \sim 620$ eV and Eq.(2) then gives $L(E_m) \sim$ 0.15 nm. In the present context however, this value may be an overestimate because the number of secondaries eventually created as the result a primary energy loss E is more closely related to the stopping power $E(d\sigma/dE)$. As a result, the appropriate average of $E_m$ will be somewhat larger and $L(E_m)$ smaller than 0.15 nm for a uranium atom.

More precise conclusions should be possible by calculating the point-spread function for inelastic scattering [30]. An estimate based on the Lenz-model angular distribution (with $E_m \sim 620$ eV) suggests that the FWHM of the PSF is about 0.04 nm, with about 40% of the inelastic intensity lying within this diameter. Combining this FWHM in quadrature with the estimated probe diameter (0.075 nm), the expected U-atom diameter in the SE image is 0.085 nm. This result would explain why the average U-atom diameter measured in the SE images was not more than 0.1 nm. In general, it does not seem difficult to understand why atomic-scale SE imaging is possible in the case of high-Z and medium-Z elements.

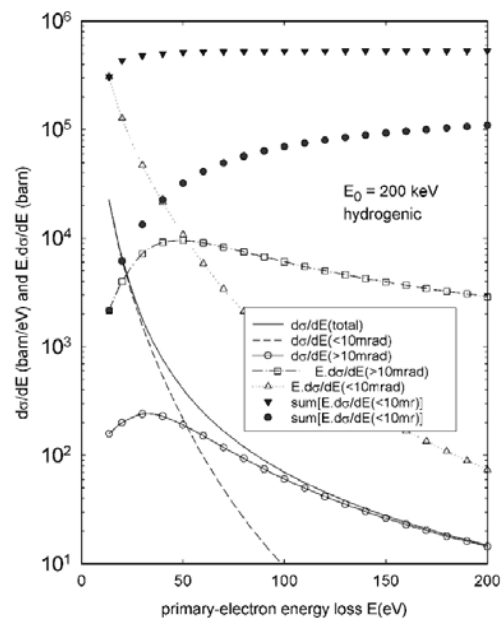

Fig.15: calculated cross section and stopping power (per electron) for excitation of hydrogenic electrons, as a function of energy loss E and for scattering angles less than and greater than 10 mrad. The solid data points show the energy-integrated stopping powers for these two angular regimes.

For elements of low atomic number, the situation is less straightforward. If we take $Z = 6$ and $E_m \sim 6.75Z \sim 40$ eV, Eq.(2) gives $L(E_m) \sim 1.2$ nm, whereas the observed spacing of graphite lattice fringes is $d \sim 0.34$ nm. Using Bragg's law, the scattering angle needed to observe these fringes in a TEM image is $\theta = \lambda/d = 7.4$ mrad. Because delocalization is directly linked to the angular distribution of inelastic scattering, we can assume that scattering angles in excess of this value are needed to achieve $L(E_m)$ low enough to provide fringes with good contrast. We must therefore ask how much of the inelastic scattering lies *above* this angle.

A simple atomic model [31] can provide us with an estimate of the fraction F of scattering (summed over all energy loss) that lies *below* an angle $\beta$:

$$F = \frac{ln\left(\frac{(1+(\beta/\theta_E)^2)(1+(\theta_0/\theta_E)^2)}{1+(\beta/\theta_E)^2+(\theta_0/\theta_E)^2}\right)}{ln(\theta_0^2/\theta_E^2)} \qquad (3)$$

where $\theta_0 \sim \lambda Z^{1/3}/(2\pi a_0)$ acts as a soft cutoff to the Lorentzian angular distribution and $\theta_E \sim 0.5 E_m/E_0$ (non-relativistically). For $\beta = 7.4$ mrad, $Z = 6$ and $E_0 = 200$ keV, $\theta_0 \sim 14$ mrad, $\theta_E \sim 0.12$ mrad and $F = 0.84$, meaning that 16% of the scattering lies at *higher* angles and would correspond to spatial frequencies larger than 0.34 nm$^{-1}$. Therefore we could expect to see evidence of carbon lattice fringes in the SE image, but with reduced contrast because of the background provided by the lower-angle inelastic scattering. This prediction appears consistent with our experimental data, where the maximum contrast in the SE images of carbon is of the order of 10%.

The results of a slightly more sophisticated calculation are shown in Fig.15. Here we use a hydrogenic model to estimate the energy-differential cross section and stopping power (integrated over all scattering angles) for the excitation of outer-shell electrons in carbon, assuming a binding energy of 13.6 eV (the σ-band width is 16 eV and the work function 9 eV in graphite). For large energy transfers (> 50 eV), the exact form of the initial-state wave functions should be relatively unimportant; the angular distribution of scattering becomes concentrated in a Bethe ridge centered around a scattering angle $(E/E_0)^{1/2}$ whose width is comparable to the binding energy [32]. The solid data points in Fig.15 show that if we include primary-electron collisions up to high energy loss, the stopping power for the large-angle (> 10 mrad) collisions reaches about 5% of that for the low-angle (< 10 mrad) collisions. Because stopping power is here a measure of how many secondary collisions can be created for a given range of energy loss and 10 mrad corresponds to a 200keV lattice resolution of 0.25 nm, we can estimate the contrast of 0.34nm lattice fringes in the SE image to be in the range 5 – 10%, which is consistent with our observations.

If the ultimate resolution of SE images is indeed determined by the delocalization of inelastic scattering, this resolution should improve somewhat as the incident-electron energy is reduced, since the product $\lambda E_0^{3/4}$ in Eq.(1) falls. Figure 16 shows the change in delocalization length with incident energy and suggests a 30% improvement in resolution at 60 keV and a factor of 2 at 30 keV. It remains to be seen whether the spherical and chromatic aberration of a probe-forming objective lens can be corrected to a sufficient degree (probe size ~ 0.1 nm) to realize these improvements. Nevertheless, there seems to be no fundamental reason why atomic resolution in SE images could not be obtained at the accelerating voltages currently used in an SEM. The extent of backscattering will be increased at lower beam energy, reducing the image contrast, but should remain weak for specimens that are thin enough to produce STEM images.

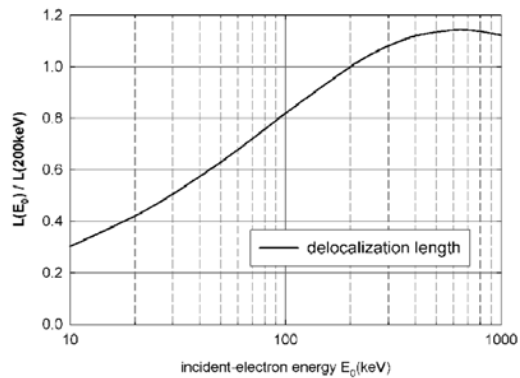

Fig.16: Delocalization length as a function of primary-electron energy, normalized to $E_0$ = 200 keV

## 5. Conclusions

We have demonstrated that using an aberration corrected scanning electron microscope with a subnanometer probe, it is possible to achieve atomic resolution using secondary electrons (SE) for a wide range of elements from uranium to carbon. The contrast of the SE images is relatively insensitive to sample thickness, crystal tilt, but sensitive to defocus, compared with its counterpart of annular dark-field (ADF) imaging using transmitted electrons. Bias experiments suggest majority of the SE have low energies (<20eV), thus are highly surface sensitive with only a few nm escape length. Some of the high-energy SE, on the other hand, can be emitted beneath a-few-nm-thick carbon amorphous layer covered on a crystal surface. A mechanism is proposed, based on the processes of SE generation, transport and emission across a surface-potential barrier (work function). The secondary electrons responsible for atomic-scale resolution in a SE image are generated by inelastic scattering events with large momentum transfer, which give rise to a sharp central peak in the point-spread function for SE generation.

The additional information obtainable by simultaneous SE and ADF imaging justifies further study including image calculations to understand not only the SE signal generation process, but also the information which can be obtained from it. One clear outcome thus far is the importance of specimen preparation of clean surfaces in gaining interpretable results.

## Acknowledgements

We would like to thank A. Howie for stimulating discussions and Y. Suzuki and K. Nakamura for technical assistance. The work was supported by the U.S. Department of Energy, Office of Basic Energy Science, under contract number DE-AC02-98CH10886.

## References

* corresponding author: zhu@bnl.gov
[1] Y. Zhu, H. Inada, K. Nakamura, and J. Wall, Nature Materials, 8 (2009) 808.
[2] D.C Joy, Nature Materials, 8 (2009) 776.

[3] Y. Zhu, and J. Wall, Aberration-corrected electron microscopes at Brookhaven National Laboratory", Book chapter in Aberration-corrected Electron Microscopy, A thematic volume of Advances in Imaging & Electron Physics, Ed., P.W.Hawkes, Elsevier/Academic Press, p.p. 481-523 (2008).
[4] H. Inada, L. Wu, J. Wall, D. Su, and Y. Zhu, J. Electron Microscopy, 58 (2009) 111.
[5] J.I. Goldstein, D.E. Newbury, P. Echlin, D.C. Joy, A.D. Romig, C.E. Lyman, C. Fiori, E. Lifshin, Scanning Electron Microscopy and X-ray Microanalysis, Plenum Press, New York, 1992.
[6] P.E. Batson, N. Dellby, and O.L. Krivanek, Nature 418, (2002) 617.
[7] A. Howie, private communication.
[8] J. Ciston, A. Subramanian, and L.D. Marks, Phys. Rev B 79 (2009) 085421
[9] R.E. Egerton, Electron Energy-Loss Spectroscopy in the Electron Microscope, Plenum Press, 1996, New York, p.302
[10] S.E. Maccagnano-Zacher, K.A. Mkhoyan, E.J. Kirkland and J. Silcox, Ultramicroscopy, 108 (2008) 718.
[11] K.A. Mkhoyan, S.E. Maccagnano-Zacher, E.J. Kirkland and J. Silcox, Ultramicroscopy, 108 (2008) 791.
[12] M. Inokuti, J.L. Dehmer, T. Baer and D.D. Hanson, Phys. Rev A 23 (1981) 23.
[13] Z.H. Yu, D.A. Muller, and J. Silcox, J.of Appl. Phys. 95 (2004) 3362.
[14] J.M. LeBeau, S.D. Findlay, L.J. Allen, and S. Stemmer, Phys. Rev. Lett. 100 (2008) 206101.
[15] D. C. Joy, Monte Carlo Modeling for Electron Microscopy and Microanalysis (Oxford University Press, New York, 1995).
[16] J.A. Venablers and J. Liu, Journal of Electron Spectroscopy and Related Phenomena 143 (2005) 205.
[17] J. Liu and J.M. Cowley, Scanning Microsc. 2, (1988) 65.
[18]  J. Liu and J. M. Cowley, Scanning Microsc. 2 (1988) 1957.
[19]  F.J. Pijper and P. Kruit, Phys. Rev. B,  44, (1991) 9192
[20] H. Mullejans 1, A.L. Bleloch, A. Howie and M. Tomita, Ultramicroscopy 52 (1993)360.
[21] J. Drucker, M.R. Scheinfein, J. Liu, and J.K. Weiss, J. Appl. Phys., 74 (1993) 7329.
[22] Darji and A. Howie,  Micron 28 (1997) 95.
[23] A. Zobelli, A. Gloter, C.P. Ewels, G. Seifert, and C. Colliex, Phys. Rev. B 75 (2007) 245402.
[24] M.R.Scheinfein, J. Drucker and J.K. Weiss, Phys. Rev. B 47 (1993)  4068,
[25] J. Drucker and M.R. Scheinfein, Phys. Rev. B (1993) 15973.
[26] L. Reimer, *Scanning Electron Microscopy*, second edition (Springer, New York), 1998.
[27] A. Howie, J. Microsc. 180 (1995) 192.
[28] L.J. Allen, S. D. Findlay, A. R. Lupini, M. P. Oxley, and S. J. Pennycook, Phys. Rev. Lett. 91, (2003) 105503.
[29] M. P. Oxley and S.J. Pennycook,  Micron 39, (2008) 676.
[30] R.F. Egerton, Ultramicroscopy **107**, (2007) 575.
[31] F. Lenz, Z. Naturforsh. 9A (1954) 185.
[32] M. Inokuti. Rev. Mod. Phys. **43** (1971) 297.